# A tunable plasmonic refractive index sensor with nanoring-strip graphene arrays

Chunlian Cen[1,2], Hang Lin[1,2], Cuiping Liang[1,2], Jing Huang[1,2], Xifang Chen[1,2], Yong Yi[1,2], Yongjian Tang[1,2], Zao Yi [1,2]*, Xin Ye[3*], Jiangwei Liu[4*], Shuyuan Xiao[5]

[1]*Joint Laboratory for Extreme Conditions Matter Properties, Southwest University of Science and Technology, Mianyang 621010, China*

[2]*Sichuan Civil-Military Integration Institute, Mianyang 621010, China*

[3]*Research Center of Laser Fusion, China Academy of Engineering Physics, Mianyang 621010, China*

[4] *School of Energy Science and Engineering, Central South University, Changsha 410083, China*

[5]*Wuhan National Laboratory for Optoelectronics, Huazhong University of Science and Technology, Wuhan 430074, China*

∗ Correspondence should be addressed to Zao Yi, Xin Ye, Jiangwei Liu

Tel: 86-0816-2480830; Fax: 86-0816-2480830

E-mail address: yizaomy@163.com; yexin@caep.cn; jiangweiliu@csu.edu.cn

**Abstract:**

In this paper, a tunable plasmonic refractive index sensor with nanoring-strip graphene arrays is numerically investigated by the finite difference time domain (FDTD) method. The simulation results exhibit that by changing the sensing medium refractive index $n_{med}$ of the structure, the sensing range of the system is large. By changing the doping level $n_g$, we noticed that the transmission characteristics can be adjusted flexibly. The resonance wavelength remains entirely the same and the transmission dip enhancement over a big range of incidence angles [0º,45º] for both TM and TE polarizations, which indicates that the resonance of the graphene nanoring-strip arrays is insensitive to angle polarization. The above results are undoubtedly a new way to realize various tunable plasmon devices, and may have a great application prospect in biosensing, detection and imaging.

## 1. Introduction

Surface plasmon resonance (SPR) on the metal/dielectric interface collective oscillation of charged carriers on sub-wavelength scales provides a combination of electronics and photonics, which makes for a wide range of applications [1,2]. Surface plasmon resonance properties depend on the nanostructure of cell plasmon geometry, size, composition and optical polarization [3,4]. This particular feature is widely used in the field of biological and chemical sensing [3-5]. In addition, a number of applications have been reported, including ultra-fast transistor photodetectors [6-8], optical modulators [9,10], light emitters [11,12], optoelectronic devices [13], biosensors [14,15] and transparent solar cells [16].

Graphene a single layer of carbon atoms are arranged in a plane with a honeycomb lattice, and when interacting with the incident light, graphene itself behaves as a metal, supporting the application of the SPR in mid-infrared (mid-infrared) and terahertz (THz) regions [17-20]. What's more interesting is that the surface conductivity of graphene can be flexible with electric field, magnetic field and chemical doping [21,22]. Because of its peculiar electrical and optical properties [23,24], there are broad application prospects in the fields of optoelectronic such as transparent electrodes [25-27], optical modulators [28,29], and photodetectors [30-32].

In this paper, we present a tunable plasmonic refractive index sensor with nanoring-strip graphene arrays. Transmission spectra depends on the sensing medium refractive index and geometrical parameters of the structure, are simulated by the FDTD method and extensively studied. By changing the sensing medium refractive index $n_{med}$ and geometrical parameters of the structure, we found that the transmission spectra of the system has a large sensing range. When plasmon resonance can restrict the nanoring-strip edges of electromagnetic field, different doping level nanoring-strip for electrically tunable spectrum detection provides interesting applications. The proposed sensor structure in biosensing, detection and imaging has potential application prospects.

2. **The geometric structure and numerical model**

We presented a simple nanoring-strip structure, as shown in figure 1. It consists of a nanoring-strip graphene arrays placed between a substrate ($n_{sub}$) and a sensing medium ($n_{med}$). $P$ is the fixed period of the arrays. The width of the nanoring is $W_1$, the width of the strip is $W_2$, the length of the strip is $L$ and thickness of graphene is $t$.

The surface conductivity of graphene can be obtained through the Kubo formula [33-35], including the interband and intraband transition contributions. The frequency range that we care about is obviously less than $2E_F$ and $\hbar\omega Oph \approx 0.2 ev$ [33], where $E_F$ is the graphene Fermi energy and ℏωOph is the phonon loss threshold. In the lower THz range, the interband transition and optical phonon emission contributions can be safely ignored. In addition, during the whole simulation calculation, the temperature $T$ was set to 300 K, and the condition of $E_F \gg K_B T$ can be satisfied for the doping levels of graphene under consideration. To sum up, according to Pauli exclusion principle, the conductivity of graphene can be approximated as an intraband Drude-like expression:

$$\sigma(\omega) = \frac{e^2 E_F}{\pi \hbar^2} \frac{i}{\omega + i/\tau} \quad (1)$$

here $\omega$ is the angular frequency of the incident electromagnetic wave, $e$ is the charge of an electron, $\hbar = h/2\pi$ is the reduced Planck's constant and $\tau$ is the relaxation time. $\tau$ and $E_F$ can be written as $\tau = \mu \hbar \sqrt{\pi |n_g|}/ev_F$ and $E_F = \hbar v_F \sqrt{\pi |n_g|}$. Where $n_g$ is the doping level of graphene, $v_F$ is the graphene Fermi velocity and $\mu \approx 10000 \text{cm}^2 \text{ V}^{-1} \text{ s}^{-1}$ is the measured dc mobility [36], similar to that used in [33,34,37,38].

In our research, we have calculated the graphene nanoring-strip spectral responses using the FDTD method. The perfectly matched layer (PML) conditions are used in the z direction and the periodical boundary conditions are utilized in the x and y directions.

3. **Simulation results and discussions**

In our work, we studied the transmission spectra of nanoring-strip (length $L$ = 180 nm, nanoring width is $W_1$ = 30 nm, strip width is $W_2$ = 30 nm and graphene thickness is $t$ = 1 nm, respectively). Through FDTD simulation calculations, we can obtain the transmission spectra of the nanoring-strip, as shown in Fig. 2(A). We can find that the nanoring-strip has an obvious transmission dip. This phenomenon can be understood by the electric field diagram shown in Figure 2(B). We find that the electric field of the nanoring-strip is mainly distributed on the nanoring arm edge of the structure and has stronger field enhancement. The transverse electric dipole resonance causes the local sympathetic vibration to be enhanced, and this strong resonance effectively captures the light energy.

Firstly, we studied the geometric structure of graphene ring-nanostrip. Other parameters are unchanged ( $W_1 = W_2 = 30$ nm, $P = 300$ nm, $n_g = 3 \times 10^{13}$ cm$^{-2}$, $t = 1$ nm and $n_{med} = 1.0$). When we changed the length $L$ from 140 nm to 220 nm, the transmission dip redshifts from 17.73 to 28.15 μm, as shown in figure 3(A). This physical mechanism can be understood by the electric field distribution diagram, as shown in figure 3(B). As the long axis increases, the gap between the adjacent nanoring-strip graphene arrays decreases, resulting in increased coupling between them. The increase of coupling leads to the red shift of the transmission dip.

In order to study the sensing properties of graphene nanoring-strip system, FDTD was used to calculate the transmission spectra of the arrays and its resonance wavelength under different refractive index of the surrounding sensing medium. In the simulation calculation, the doping level of graphene is $n_g = 3 \times 10^{13}$ cm$^{-1}$, $W_1 = W_2 = 30$ nm, $L = 180$ nm and $P = 300$ nm, respectively. It is clear in figure 4(A) that the transmission spectra of mode A and mode B have changed significantly in the refractive index of different sensing medium $n_{med}$. Clearly seen in figure 4(A) under different refractive index of sensing medium $n_{med}$, obvious changes in the transmission spectra of mode A and mode B have taken placed, at the same time, notice the sensing range of the system is very big. In order to quantify the refractive index of the sensor of the proposed performance, we calculated the full width at half maximum (FWHM) and figure of merit (FOM) of mode B for the refractive indices of different sensing medium, as show in figure 4(B) the red curve (FWHM) and the black curve (FOM). We noticed that the resonance wavelength shifts in refractive index change linearly, namely, $\Delta\lambda = m\Delta n$. Here, $\Delta\lambda$ is resonance wavelength, $m$ is the resonance wavelength shift per refractive index unit (RIU) change and has units of meters per RIU, $\Delta n$ is range of the refractive index of sensing medium $n_{med}$, respectively. We can get the sensitivity $S = m = \partial\lambda/\partial n$ by using the above formula. The sensitivity of mode A and mode B are 2.97 μm/RUI and 5.20 μm/RUI, respectively. The sensitivity of mode B is higher than that of mode A. For mode B, a trend that increases linearly with increasing refractive index of the sensing medium $n_{med}$. This is due to increased $n_{med}$ can increase the nanoring-strip the damping dipole mode, which result in an increase of the FWHM of transmission curve. According to the formula [39] $FOM = \dfrac{m}{FWHW}$, the FOM decrease with the refractive index of the sensing medium $n_{med}$.

To analyze the tunable properties of the ring structure we have proposed, when $L$ = 180 nm, $W_1$ = $W_2$ = 30 nm and $P$ = 300 nm, we calculate the transmission spectra by changing the doping level $n_g$, as shown in figure 5(A). We found that with the increase of doping level $n_g$, graphene nanoring-strip had stronger transmission dip. Therefore, by changing doping level, the transmission spectra of graphene nanoring-strip can be effectively changed, indicating that the proposed structure has good tunable characteristic. This will have wide application prospects in the field of sensors. Because the doping level $n_g$ and the duty ratio arrays are fixed for different $W_1$ values, the corresponding transmission spectra show almost the same minimum value, in figure 5(B). In figure 5(B), when the width varying from 30 nm to 50 nm, the transmission spectra also has the very high sensitivity, but its resonance wavelength change is very small. Through compared with figure 3, we found that the length of the graphene nanoring-strip has a wider resonance wavelength.

The situations of the transmission in graphene nanoring-strip on the incidence angle and polarization are also researched. In figure 6(A) and (B), the resonance wavelength remains entirely the same and the transmission dip enhancement over a big range of incidence angles [0º,45º] for both TM and TE polarizations. There are two main reasons for the good operation angle polarization tolerance. In the first place, the graphene nanoring-strip has a high rotational symmetry. In addition, the transmission dip enhancement here is derived from the strongly localized surface plasmonic resonance.

In figure 6(A) and (B) the resonance wavelength is very stable for all incident angles, indicating that the graphene nanoring-strip arrays is insensitive. Consequently, the graphene nanoring-strip arrays can be applied to angle-independent devices.

**4. Conclusions**

In a word, a tunable plasmonic refractive index sensor with nanoring-strip graphene arrays has been designed and theoretically demonstrated. Through the simulation calculations, we found that the transmission spectra of nanoring-strip has a high sensitivity. By changing the sensing medium refractive index $n_{med}$ of the nanoring-strip structure, the system has a large sensing range. When changing doping level $n_g$, we noticed that the transmission characteristics of nanoring-strip can be tunabled flexibly. The results also show that the graphene nanoring-strip arrays is insensitive to different polarization modes (TM or TE). With nanoring-strip graphene arrays have a good tunability and a high sensitivity, we believed that this kind of sensor has a great application prospect in biosensing, detection and imaging.

**Acknowledgements**

The work is supported by the National Natural Science Foundation of China (No. 51606158; 11604311; 61705204; 21506257), the Funded by Longshan academic talent research supporting program of SWUST (No. 17LZX452).

**Figure captions:**

**Figure 1.** The schematic diagram of geometric structure is designed as follows: nanoring-strip graphene arrays with period *P*, length *L*, nanoring width $W_1$, strip width $W_2$ and graphene thickness *t*. The arrays are supported on a substrate ($n_{sub}$) and a sensing medium ($n_{med}$).

**Figure 2.** (A) The transmission spectra of the three structures of nanoring-strip (length *L = 180 nm*, nanoring width $W_1$ = 30 nm and strip width $W_2$ = 30 nm). (B) The electric field distribution of nanoring-strip structure.

**Figure 3.** The transmission spectra of graphene with different length (L). Other parameters are unchanged ( $W_1$ = $W_2$ = 30 nm, *P* = 300 nm, $n_g$ = 3 x $10^{13}$ cm$^{-2}$ and $n_{med}$ = 1.0).

**Figure 4.** (A) The transmission spectra of graphene with different sensing medium ($n_{med}$). (B) FWHM and FOM of mode B for the refractive indices of different sensing medium. Other parameters are unchanged (*L* = 180 nm, *P* = 300 nm, $W_1$ = $W_2$ = 30 nm, $n_g$ = 3 x $10^{13}$ cm$^{-2}$).

**Figure 5.** (A)The transmission spectra of graphene with different doping level ($n_g$). (B) The transmission spectra of graphene with different ring width ($W_1$). Other parameters are unchanged (*L* = 180 nm, $W_2$ = 30 nm, *P* = 300 nm and $n_{med}$ = 1.0).

**Figure 6.** The simulated angular dispersions of the transmission in graphene nanoring-strip with the doping level of $n_g$ = 3 x $10^{13}$ cm$^{-2}$ for (A) TE and (B) TM.

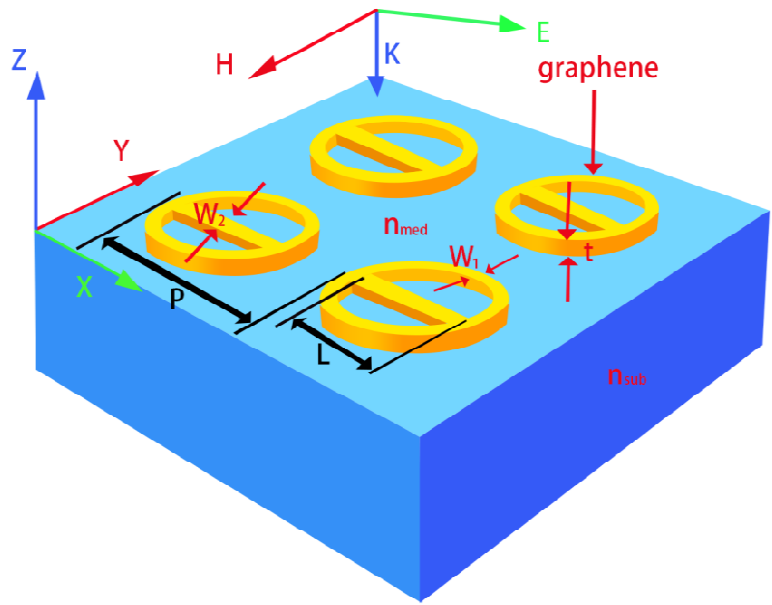

**Figure 1.**

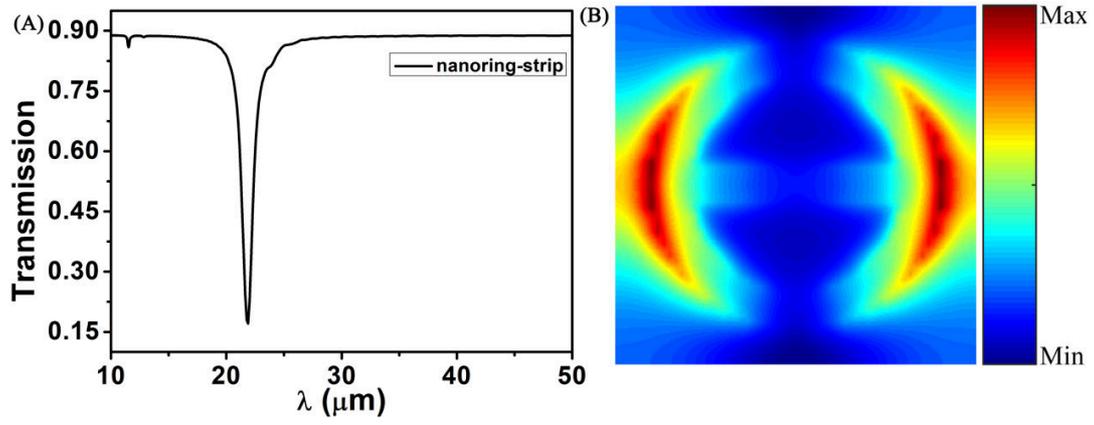

**Figure 2.**

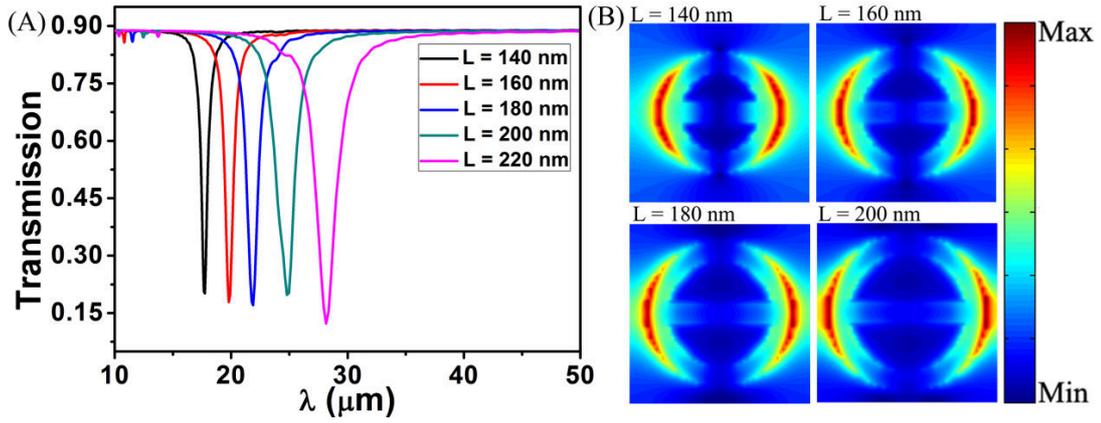

**Figure 3.**

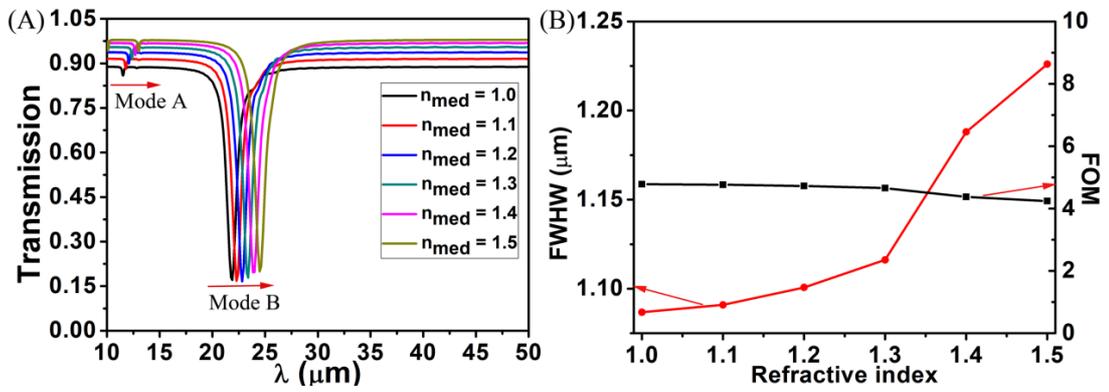

**Figure 4.**

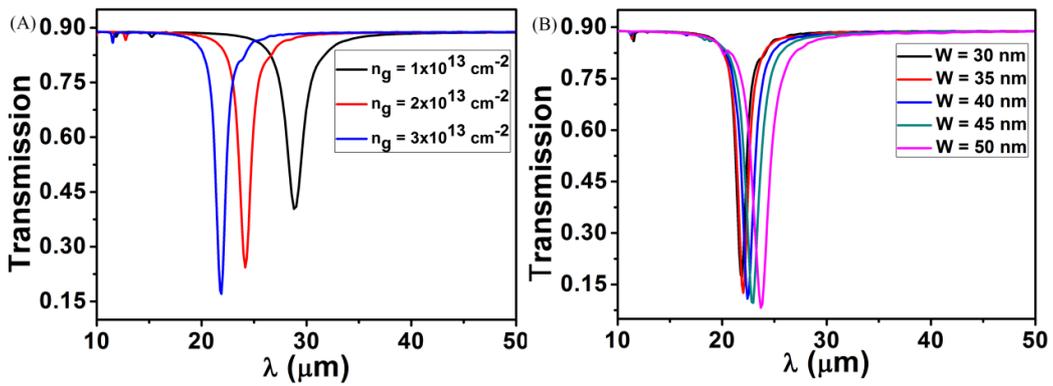

**Figure 5.**

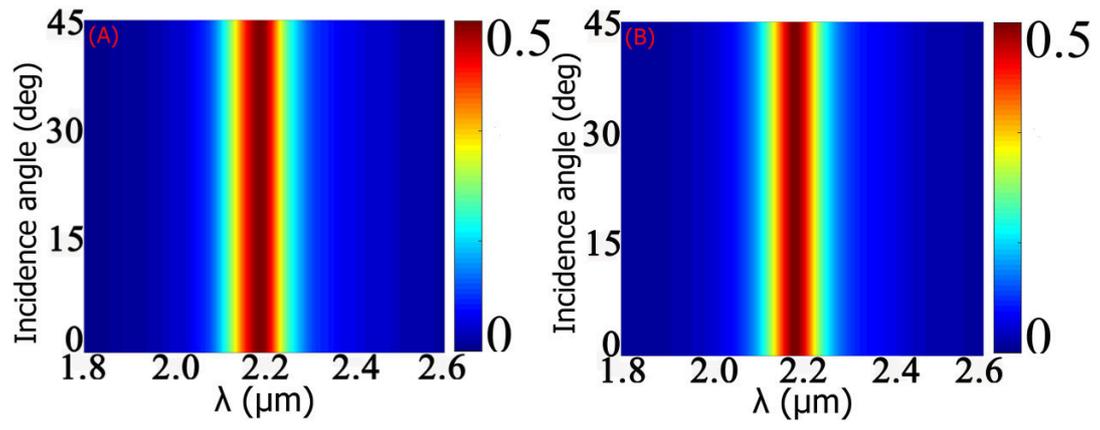

Figure 6.